\documentclass[12pt]{article}
\usepackage{epsfig}
\begin{document} 

\vspace{2em}
\hspace*{-0.5cm}
\begin{center}
{\Large {\bf TeV NEUTRINOS IN A DENSE MEDIUM}} 

\vspace*{2cm}

{\large A. Nicolaidis$^{a}$,
G.~Tsirigoti$^{a}$ and J. Hansson$^{b}$}\footnote{
e-mail addresses : nicolaid@ccf.auth.gr, georgia@ccf.auth.gr,
hansson@mt.luth.se} 

$^{(a)}$~Department of Theoretical
Physics, University of Thessaloniki, GR-54006 Thessaloniki, Greece\\ 
$^{(b)}$~Department of Physics, Lule\aa University of Technology,
SE-97187  Lule\aa, Sweden\\

\vspace*{5cm}

{\bf Abstract}\hspace{2.2cm}\null

\end{center}

The dispersion relation of energetic (few TeV) neutrinos
traversing a medium is studied. We use the real time formalism
of thermal field theory and we include the effects from the
propagator of the W gauge boson. We consider then the MSW
oscillations for cosmic neutrinos traversing the Earth, adopting
for the neutrino parameters values suggested by the LSND
results. It is found that the $\nu_\mu$ flux, for neutrinos
passing through the center of the Earth, will appear reduced by
15\% for energies around 10 TeV.
%
%
\vspace{2em}

\newpage

\section {Introduction}

The quest to learn whether neutrinos are massive has been long
and arduous. In the recent years there are mounting
astrophysical and laboratory data suggesting that neutrinos
oscillate from one flavor to another,which can only happen if
they have non-zero mass. The SuperKamiokande experiment on
atmospheric neutrinos\cite{fuku}, solar neutrino
experiments\cite{hata} and LSND accelerator
experiment\cite{ath}, provide three distinct scales of neutrino
mass-squared differences:\\
\begin{eqnarray}
\Delta m^{2}_{atm} \sim 5 \times 10^{-3} {\rm eV}^2\;\nonumber\\[1em]
\Delta m^{2}_{sun} \sim 10^{-5} {\rm eV}^2\ \ \\[1em]
\Delta m^{2}_{\scriptscriptstyle LSND} \sim 10~ {\rm eV}^2 \;\nonumber
\end{eqnarray}

To accomodate all data a mixing scheme with four massive
neutrinos, three active $(\nu_{e}, \nu_{\mu}, \nu_{\tau})$ and
one sterile $(\nu_{s})$ is required. The mass pattern contains
two pairs, each consisting of two nearly degenerate states,
separated by the `LSND gap' of a few eV\cite{bil}. A natural
assignment would have $(\nu_{\mu}, \nu_{\tau})$ with a mass of
few eV each and a mass-difference $\Delta m^{2}_{atm}$
accounting for the atmospheric neutrino oscillations, while
$\nu_{e}$ and $\nu_{s}$, much lighter and with a splitting
$\Delta m^{2}_{sun}$, provide the MSW solution for the solar
neutrinos.

Few - eV mass neutrinos play a cosmological role as a hot dark
matter (HDM) component in the popular cold+hot dark matter
(CHDM) model\cite{dm}, providing less structure on small scales.
In addition, neutrinos with eV masses mimic extra radiation at
the epoch of matter-radiation equality and affect the cosmic
microwave background (CMB) anisotropy\cite{ma}. Sterile
background neutrinos may generate large neutrino asymmetries and
influence the big bang nucleosynthesis (BBN)\cite{bbn}. Massive
relic neutrinos participate in gravitational clustering around
galaxies. Recent calculations\cite{niclds} indicate that
gravitational clustering provides locally an increase in the
neutrino density by a factor $10^5$, over the uniform density of
the big bang cosmology. It has been suggested\cite{fwwy} that
the highest energy cosmic rays result from the annihilation of
extremely high energy cosmic neutrinos on the background of
gravitationally clustered relic neutrinos. From all the above it
is clear that the issue of neutrinos with a mass of few eV is an
interesting one and deserves further study. We suggest in this
paper another approach to explore this issue.

Solar neutrinos, in the MeV energy range, undergo resonant
conversion with $\Delta m^{2}_{sun}~\sim 10^{-5}~{\rm eV}^2$. We
infer that the range $\Delta m^{2}_{\scriptscriptstyle LSND}
\sim 10~ {\rm eV}^2$ can be studied more appropriately, via MSW
oscillations, using neutrinos in the energy range of few TeV.
The outer space provides powerful radiation sources, like Active
Galactic Nuclei(AGN). AGN are the central regions of certain
galaxies in which the emission of radiation can rival or even
surpass the total power output of the entire galaxy by as much
as a thousand fold. The source that powers AGN is believed to be
gravity, i.e. matter accretion
into a supermassive black hole located at the center of the
galaxy. Accelerated protons may interact with matter or
radiation in the AGN to produce pions whose decay products
include neutrinos. It
is anticipated that AGN could be the most luminous high energy
neutrino sources in the universe and the diffuse isotropic
neutrino flux from all AGN has been estimated\cite{agn}. Also
Gamma-Ray Burst (GRB) sources have been proposed as generators
of high energy neutrinos\cite{agn,wax}.     

Cosmic high energy muon neutrinos (${\rm E_\nu > 1~TeV}$) can be
observed by neutrino telescopes\cite{gais,amanda} by
detecting the long range muons produced in charged current muon
neutrino-nucleon interactions. The effective detector volume is
enhanced in proportion to the range of the produced muon
(typically several kilometers). To reduce background, at the
detection site one looks for upward moving muons induced by
neutrinos traversing the Earth. We study then neutrino matter
oscillations inside the Earth ($\nu_\mu\leftrightarrow\nu_e$)
for neutrino energies above 1 TeV and with mass-squared
difference and vacuum mixing angle for the neutrinos as
suggested by the LSND results.

The usual MSW analysis has previously been applied to low energy
neutrinos, where the Fermi model is an adequate description.
Since we are 
interested in energetic neutrinos (${\rm E_\nu > 1~TeV}$), the
gauge-boson propagator effects might be important. In the next
section we study the dispersion relation for energetic neutrinos
in a medium, using the real-time formalism of finite-temperature
field theory. In section 3 we apply our results to neutrino
matter oscillations inside the Earth and present our
conclusions. 

\section{Dispersion relation for neutrinos}

At finite density and temperature the properties of particles
deviate from their vacuum values. This applies in particular to
the dispersion relation which governs their plane-wave
propagation. For neutrinos traversing a "flavor birefringent"
medium, the effective mixing angle can become large, inducing a
complete reversal of the flavor content of the neutrino state
(MSW effect\cite{wolf}). The real-time formalism of thermal
field theory\cite{das} is well suited to analyze such a problem.
The real-time propagator consists of a sum of two parts - one
corresponding to the propagator at zero temperature and the
other corresponding to a temperature dependent part. The fermion
propagator is given by
\begin{eqnarray}
S_f(p) = (\rlap / p + m) \bigl[
\frac{1}{p^2-m^2+i\varepsilon} + i\Gamma (p)\bigr]\ \ \\[1em]
\Gamma (p) = 2\pi~ \delta (p^2-m^2)~ \theta (p_o)~ \eta_f (\vert
p_o \vert) \ \ \\[1em]
\eta_f (\vert p_o \vert ) = \frac{1}{e^{\beta (\vert p_o \vert -
\mu )} + 1}
\end{eqnarray}
where $\eta_f$ is the occupation number for the background
fermions at temperature $T = \frac{1}{\beta}$ and chemical
potential $\mu$. While the zero temperature part of the
propagator can be thought of in the conventional manner as
representing the exchange of a virtual particle, the temperature
dependent part, see eq.(3), represents an on-shell contribution.
In a plasma of particles, there is a distribution of real
particles and these real particles participate in the scattering
process in addition to the exchange of virtual particles.

Neutrinos propagating in a medium acquire an effective mass.
Denoting by $\Sigma$ the modified self-energy, the propagation
of a neutrino in a medium is governed by the Dirac equation
\begin{equation}
\bigl [ \rlap /k - \Sigma (k) \bigl ] \psi = 0
\end{equation}
At one-loop level $\Sigma$ has the general form\cite{nora}
\begin{equation}
\Sigma = m - (\alpha \rlap /k + b \rlap /u)\frac{1}{2}
(1-\gamma_5) 
\end{equation}
where m is vacuum neutrino mass, k is the neutrino four-momentum
and u is the four-velocity of the medium. The coefficients
$\alpha$, b are functions of the scalar quantities $k^2,~u^2 =
1$ and $k\cdot u = \omega$, with $\omega$ the energy of the
neutrino in the rest frame of the medium. The dispersion
relation is given by
\begin{equation}
det( \rlap /k - \Sigma ) = 0
\end{equation}
and to lowest order in small quantities provides
\begin{equation}
\omega^2 - \vert \vec k \vert^2 = m^2 - 2 b \omega
\end{equation}
Equivalently the refraction index n $(\vert \vec k \vert = n
\omega)$ becomes
\begin{equation}
n = 1 - \frac{m^2}{2\omega^2} + \frac{b}{\omega}
\end{equation}

Neutrinos traversing the Earth interact with electrons via
neutral current and charged current interactions. While all
active neutrinos interact via the neutral current, only the
$\nu_e$ participate in the charged current interaction. The
relevant one loop contribution to the self-energy of the
neutrino is given by (see Fig. 1)

\begin{center}
\[
\epsfxsize=7cm
\epsffile[57 98 576 715]{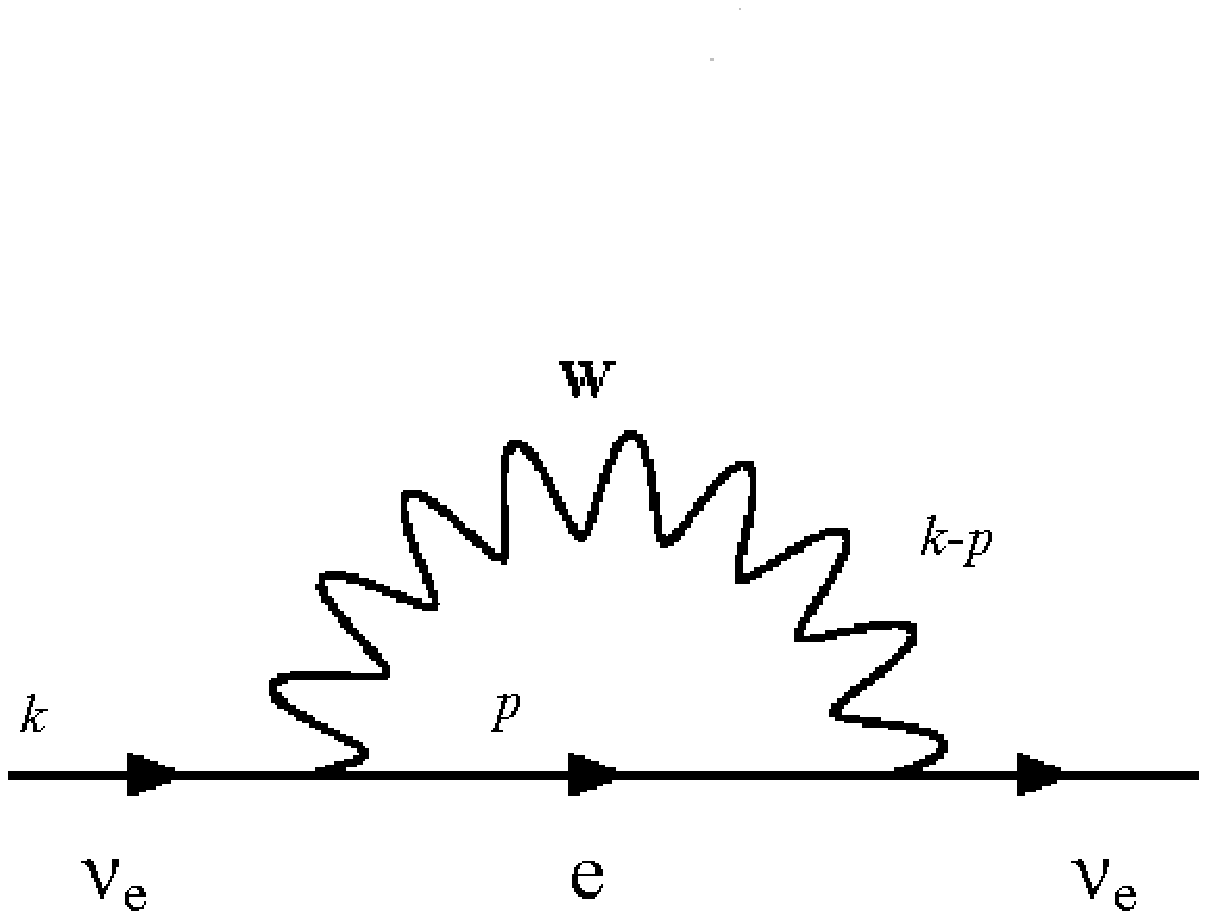}\]
\end{center}
\vspace{-6.5cm}

\begin{center}
Fig. 1 \ \ \ \ \ \ \  
\end{center}

\begin{equation}
i\Sigma = - \bigl ( \frac{ig}{2\sqrt{2}}\bigl )^2 \int \frac{d^4
p}{(2\pi)^4} \gamma^\mu (1 - \gamma_5) i (\rlap /p + m) i \Gamma
(p) D_{\mu\nu} (k-p)
\end{equation}
where g is the SU(2) coupling constant and $D_{\mu\nu}$ is the
propagator for the W gauge boson. In the unitary gauge we have 
\begin{equation}
D_{\mu\nu} (k-p) = \frac{-i}{(k-p)^2 - M_{\scriptstyle W}^2}
\bigl [ g_{\mu\nu} 
- \frac{(k-p)_\mu (k-p)_\nu}{M_{\scriptstyle W}^2} \bigl ]
\end{equation}

Usually the energies involved are small compared to
$M_{\scriptstyle W}$ and the momentum dependence in the W
propagator is ignored. Since we are interested in energetic
neutrinos (${\rm E_\nu > 1~TeV}$) we expand the propagator in
powers of $M_{\scriptstyle W}^{-2}$. We obtain 
\begin{equation}
D_{\mu\nu}(k-p) \simeq i\bigl [
\frac{g_{\mu\nu}}{M_{\scriptstyle W}^2} +
\frac{(k-p)^2 g_{\mu\nu} - (k-p)_\mu (k-p)_\nu}{M_{\scriptstyle
W}^4} \bigl ] 
\end{equation}

It is the above expression we have employed in our calculations.
Notice also that from the electron propagator we kept only the
term emanating from the medium. The first term in eq.(2),
electron propagator in the vacuum, provides the wave-function
renormalization of the neutrino in the vacuum (of no relevance
to us).

Working within the Fermi model approximation, i.e. keeping only
the first term from the W propagator, eq.(12), we obtain for the
corresponding self-energy $\Sigma_o$ (evaluated in the rest
frame of the medium)
\begin{equation}
i\Sigma_o = i G_{\scriptstyle F} \sqrt{2}~ \gamma^{\scriptstyle 0}
~\frac{1}{2} (1-\gamma_5)~ 2 \int \frac{d^3p}{(2\pi)^3} \eta_f
\end{equation}
Since the number density of the background electrons is given by
\begin{equation}
N_e = 2 \int \frac{d^3p}{(2\pi)^3} \eta_f
\end{equation}
we arrive at the following expression for $\Sigma_o$, valid in
an arbitrary frame
\begin{equation}
i\Sigma_o = i G_{\scriptstyle F} \sqrt{2}~N_e~\rlap /u
~\frac{1}{2} (1-\gamma_5)~ 
\end{equation}   
We read off then that 
\begin{equation}
b_o = -\sqrt{2} G_{\scriptstyle F} N_e
\end{equation}
in agreement with previous calculations
[16-21]. 

Including the second term from eq.(12), i.e. the correction from
the finite mass W propagator, we find for the corresponding
self-energy $\Sigma_1$ (evaluated also in the rest frame of the
medium) 
\begin{equation}
i\Sigma_1 = -i \frac{g^2}{M_{\scriptstyle W}^4} E_\nu \gamma_o
\frac{1}{2} 
(1-\gamma_5) \int \frac{d^3p}{(2\pi)^3} \eta_f (\varepsilon)
(\varepsilon + \frac{\vec p^2}{3\varepsilon}) ~+ \rlap /k ~{\rm
terms}
\end{equation}
By $\rlap /k ~{\rm terms}$ we denote all these terms
proportional to $\rlap /k$, responsible for the $\alpha$
coefficient (see eqn.(6)). Since we are interested in finding
the b coefficient we ignore these terms. $\varepsilon$ and $\vec
p$ are the energy and the momentum of the background fermions.
We consider two distinct cases: relativistic electrons (R),
where $\vec p^2 \simeq \varepsilon^2$, and non-relativistic
electrons (NR), where $\vec p^2 <<\varepsilon^2$. Then we have 
\begin{equation}
\varepsilon + \frac{\vec p^2}{3\varepsilon} \simeq c \varepsilon 
\end{equation}
with $c = \frac{4}{3}$(1) for the R(NR) case. For
$\Sigma_1$ in an arbitrary frame we obtain 
\begin{equation}
i\Sigma_1 = -\frac{ig^2}{2 M^4_{\scriptstyle W}} E_\nu N_e c
<\varepsilon> 
\rlap /u \frac{1}{2} (1-\gamma_5) + \rlap /k~ {\rm terms}
\end{equation}
with $<\varepsilon>$ the average energy of the electrons in the
medium. We extract that 
\begin{equation}
b_1 = 2\sqrt{2} c G_{\scriptstyle F} N_e \frac{E_\nu
<\varepsilon>}{M_{\scriptstyle W}^2}
\end{equation}  
Altogether
\begin{equation}
b=b_o + b_1 = -\sqrt{2} G_{\scriptstyle F} N_e\bigl [ 1-2 c
\frac{E_\nu <\varepsilon>}{M_{\scriptstyle W}^2}\bigl ]
\end{equation}
With $E_\nu$ in the TeV range and electrons in a hot plasma the
second term in the square bracket could be a significant
correction. In our case the electrons inside the Earth can be
considered as a Fermi gas and the maximum momentum
$p_{\scriptscriptstyle F}$ is fixed by the local number density.
For the actual Earth densities we find low values for
$p_{\scriptscriptstyle F}$. Therefore we obtain 
\begin{equation}
b=-\sqrt{2} G_{\scriptstyle F} N_e\bigl [ 1-2\frac{E_\nu
m_e}{M_{\scriptstyle W}^2}\bigl ] 
\end{equation}
We deduce that the correction is not significant in the Earth
and in our calculations we use the standard MSW formalism.
\section{Results and Conclusions}

Foccusing our attention into $\nu_\mu <-> \nu_e$ oscillations,
any neutrino state can be written as
\begin{equation}
\vert \nu (t)> = \alpha_e (t) \vert \nu_e> + \alpha_\mu (t)
\vert \nu_\mu>
\end{equation}
and the time evolution is given by 
\begin{equation}
i\frac{d}{dt}\left (\matrix {\alpha_e\cr\alpha_\mu\cr}\right ) =
{\large A} \left (\matrix {\alpha_e\cr\alpha_\mu\cr}\right )
\end{equation}
where
\begin{equation}
{\Large A} = \frac{1}{4p} \left (\matrix {-\Delta
cos2\theta_o+2p\gamma & 
\Delta sin2\theta_o\cr \Delta sin2\theta_o & \Delta
cos2\theta_o-2p\gamma\cr}\right )
\end{equation}  
The neutrino momentum is denoted by p, $\theta_o$ is the vacuum
mixing angle and $\Delta$ is the mass-squared difference. Matter
effects are determined by $\gamma$\cite{wolf}
\begin{equation}
\gamma = \sqrt{2} G_{\scriptstyle F} N_e
\end{equation}
with $N_e$ the density of electrons in the medium. For the
neutrino parameters we have chosen values implied by the LSND
results 
\begin{equation}
\Delta = 10 ~{\rm eV^2}, ~~sin2\theta_o = 0.07
\end{equation}
The relative population of cosmic neutrinos at the source is
$2~\nu_\mu ~:~ 1~\nu_e$. The smallness of the vacuum mixing
angle suggests that vacuum oscillations (during the journey from
the cosmic source to the Earth) do not alter the relative
proportion. However during the passage through the Earth matter
oscillations might change significantly the relative abundance
of $\nu_\mu$ and $\nu_e$. The density of the Earth is accurately
described by the two density model, where the core and the
mantle each have a separate and constant density \cite{argn}.
Defining by $\phi$ the angle between the neutrino direction and
the tangent to the surface of the Earth, we considered neutrinos
incident at $\phi = \frac{\pi}{2}$ (i.e. passing through the
center of the Earth). In the absence of any oscillations, the
muon neutrino flux exiting the Earth is identical to the initial
incident muon neutrino flux.
In the presence of oscillations, the muon neutrino flux to be
detected, is equal to the initial incident muon neutrino flux
times the factor [ $P (\nu_\mu \rightarrow \nu_\mu) +
\frac{1}{2} P (\nu_e \rightarrow \nu_\mu)$]. Fig. 2 presents
this factor as a function of the energy. At around 10 TeV and
for $\phi = \frac{\pi}{2}$, a 15\% reduction in the number of
muon neutrinos is observed. 
Notice that the observed dip corresponds to an MSW resonance.
However at resonance the oscillation length is given by
\cite{argn} 
\begin{equation}
L_{\scriptstyle R} = \frac{\pi}{\gamma tan2\theta_o}
\end{equation}
For small vacuum mixing angle $\theta_o$, $L_{\scriptstyle R}$
becomes large and exceeds the radius of the Earth. For this
reason the MSW dip is not fully developed.

Neutrino telescopes will provide a new observation window on our
cosmos and will help us probe the deepest reaches of distant
astrophysical objects. The large energies involved might also
allow us to enlarge our knowledge of particle physics.
Determining the direction of the incident neutrino, we infer the
neutrino path through the Earth. Therefore the neutrino
telescopes are suited for the study of neutrino oscillations in
the GeV-TeV energy range, and for oscillation lengths of the
order of the size of the Earth. Furthermore by comparing the
event rates of neutrinos scratching the Earth, to neutrinos
traversing the Earth, we obtain access to the neutrino
oscillation probabilities. In our work we studied the
oscillations among $\nu_\mu$ and $\nu_e$. Another possibility is
oscillations among $\nu_\mu$ and $\nu_\tau$ and the appearance
of $\nu_\tau$ at TeV energies\cite{Lpak,halzen}. This
interesting case is under study.\\[1em]
{\bf Acknowledgements :} This work was partially supported by
the EU program `Human Capital and Mobility', under contract No
CHRX-CT94-0450.    


\clearpage
\newpage
\begin{center}
\[
\epsfxsize=12cm
\epsffile[57 135 576 715]{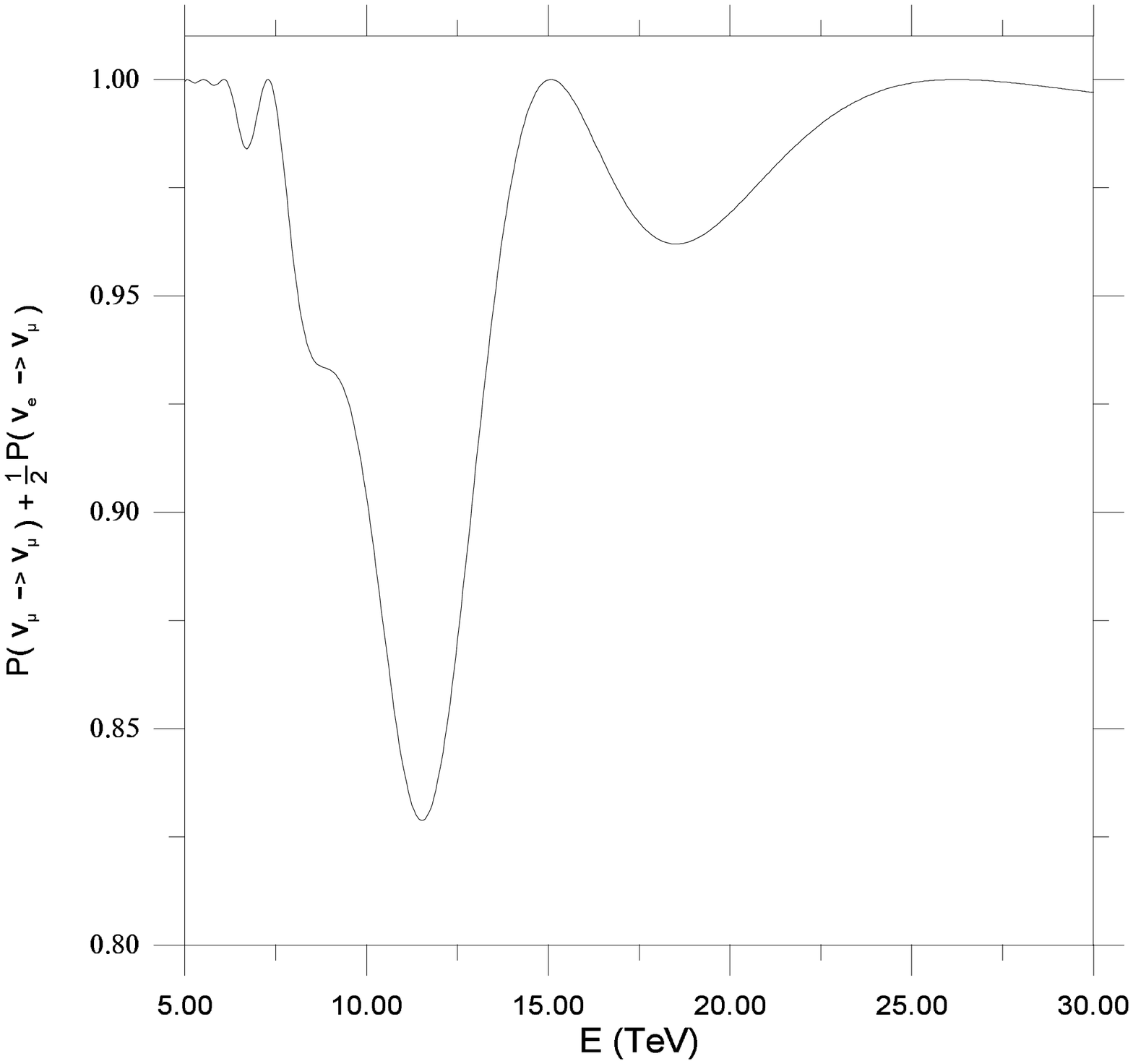}
\ \ \ \ \ \ \ \ \] 
\end{center}
\vspace{0.3cm}
\begin{center}
Fig. 2
\end{center}
\end{document}